\documentclass[%
superscriptaddress,
 amsmath,amssymb,
 aps,
 prl,twocolumn,
]{revtex4-1}

\usepackage{graphicx}% Include figure files
\usepackage{dcolumn}% Align table columns on decimal point
\usepackage{bm}% bold math
\usepackage{color} 

\begin{document}

%\title{Structural pivot  point controls electronic properties in (Fe,Zn)$_2$Mo$_3$O$_8$}% Force line breaks with \\

\title{Band-Mott mixing hybridizes the gap in Fe$_2$Mo$_3$O$_8$}

\author{K. Park}
\affiliation{These authors contributed equally to this work}
\affiliation{Department of Chemistry, University of Tennessee, Knoxville, Tennessee 37996, USA}

\author{G. L. Pascut}
\affiliation{These authors contributed equally to this work}
\affiliation{MANSiD Research Center and Faculty of Forestry, Applied Ecology Laboratory, Stefan Cel Mare University (USV), 13 University Rd, Suceava 720229, Romania}
%\affiliation{Department of Physics and Astronomy, Rutgers University, Piscataway, New Jersey 08854, USA}

\author{G. Khanal}
\affiliation{Department of Physics and Astronomy, Rutgers University, Piscataway, New Jersey 08854, USA}

\author{M. O. Yokosuk}
\affiliation{Department of Chemistry, University of Tennessee, Knoxville, Tennessee 37996, USA}

\author{Xianghan Xu}
\affiliation{Department of Physics and Astronomy, Rutgers University, Piscataway, New Jersey 08854, USA}

\author{Bin Gao}
\affiliation{Department of Physics and Astronomy, Rice University, Houston, Texas 77005, USA}
%\affiliation{Rutgers Center for Emergent Materials, Rutgers University, Piscataway, New Jersey 08854, USA}

\author{M. J. Gutmann}
\affiliation{ISIS Facility, STFC-Rutherford Appleton Laboratory, Didcot OX11 OQX, United Kingdom}

\author{A. P. Litvinchuk}
\affiliation{Texas Center for Superconductivity and Department of Physics, University of Houston, Houston, Texas
77204, USA}

\author{V. Kiryukhin}
\affiliation{Department of Physics and Astronomy, Rutgers University, Piscataway, New Jersey 08854, USA}
\affiliation{Rutgers Center for Emergent Materials, Rutgers University, Piscataway, New Jersey 08854, USA}

\author{S. -W. Cheong}
\affiliation{Department of Physics and Astronomy, Rutgers University, Piscataway, New Jersey 08854, USA}
\affiliation{Rutgers Center for Emergent Materials, Rutgers University, Piscataway, New Jersey 08854, USA}
\affiliation{Laboratory for Pohang Emergent Materials and Max Planck POSTECH Center for Complex Phase Materials, Pohang University of Science and Technology, Pohang 790-784, Korea}

\author{D. Vanderbilt}
\affiliation{Department of Physics and Astronomy, Rutgers University, Piscataway, New Jersey 08854, USA}

\author{K. Haule}
\affiliation{Department of Physics and Astronomy, Rutgers University, Piscataway, New Jersey 08854, USA}

\author{J. L. Musfeldt}
\email{musfeldt@utk.edu}
\affiliation{Department of Chemistry, University of Tennessee, Knoxville, Tennessee 37996, USA}
\affiliation{Department of Physics and Astronomy, University of Tennessee, Knoxville, Tennessee 37996, USA}

\date{\today}

\begin{abstract}

We combined optical spectroscopy and first principles electronic structure calculations to reveal 
the charge gap in the polar magnet Fe$_2$Mo$_3$O$_8$.  
Iron occupation on the octahedral site draws the gap strongly downward compared to the Zn parent compound, and subsequent
occupation of the tetrahedral site creates a narrow resonance near the Fermi energy that draws the gap downward even further. 
This resonance is a many-body effect that emanates from a flat valence band in a Mott-like state due to screening of the local moment - similar to expectations for a Zhang-Rice singlet, except that here, it appears in a semi-conductor.  We discuss the unusual hybridization in terms of  orbital occupation  and character as well as the structure-property 
relationships that can be unveiled in various metal-substituted systems (Ni, Mn, Co, Zn).

\end{abstract}

\maketitle

%%%%%%%%%%%%%%%%%%%%%%%%%%%%%%%%%%%%%%%%%%%%%%%%%%%%%%%%%%%%

\section*{Introduction}

%A simple paper explaining hybridization trends in these materials.

3$d$-containing materials are well known for strong electron correlations, narrow band widths, site- and orbital-selective states, and robust magnetism whereas  %\cite{Kotliar2006,Haule2017,Pascut2020,Lichtenstein2001,Anisimov2002,deMedici2005}, 
4- and 5$d$ systems are recognized for strong spin–orbit coupling, increased hybridization,  extended orbitals, and a tendency toward dimerization \cite{Streltsov2017}. %\cite{Birol2015,Zhang2017,Pascut2014,Radaelli2002,Georges2013} 
Combining these qualities %traits 
in mixed metal %3$d$/4$d$ and 3$d$/5$d$
materials leads to a variety of unexpected properties. 
Examples include interpenetrating sublattices with independent spin dynamics and ground states in Sr$_2$CoOsO$_6$ \cite{Morrow2013,Yan2014}, self-healing photoelectrode materials like CuRhO$_2$ \cite{Gu2014}, covalency-driven collapse of spin-orbit coupling in Ba$_5$CuIr$_3$O$_{12}$ \cite{Ye2018},  an ultra-high coercive field in Sr$_3$NiIrO$_6$ \cite{Singleton2016,ONeal2019}, magnetoelectric coupling in Co$_4$Nb$_2$O$_9$ \cite{Khanh2016}, surprising spin entropy effects across the magnetic quantum phase transition in CoNb$_2$O$_6$ \cite{Liang2015}, and nonreciprocal directional dichroism in Ni$_3$TeO$_6$ \cite{Yokosuk2020}. 
Another mixed metal system with exciting properties and curious hybridization is 
Fe$_2$Mo$_3$O$_8$ - also known as the mineral Kamiokite \cite{Inosov2018}. While magnetism and magnetoelectric coupling have been widely studied \cite{Sheckelton2012,Mourigal2014,Wang2015,Kurumaji2015,Li2017,Chen2018,Solovyev2019,Nikolaev2021}, the charge excitations are highly under-explored.

\begin{figure*}[tbh]
\begin{minipage}{7.0in}
\includegraphics[width = 7.0in]{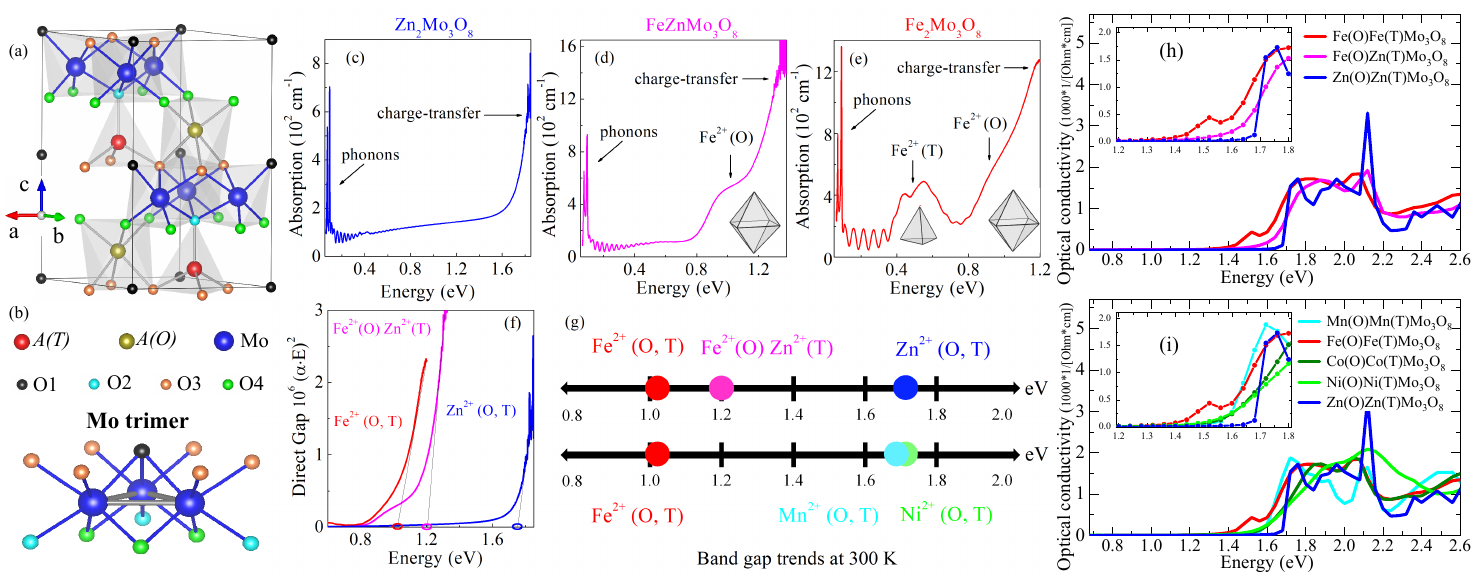}
\end{minipage}
\begin{minipage}{7.0in}
\caption{\label{FMOBandgap} 
(a) Crystal structure of the $A_2$Mo$_3$O$_8$ compounds where $A$ = Mn, Fe, Co, Ni, Zn.  $A$(T) and $A$(O) represent  ions in  trigonally-distorted tetrahedral and octahedral environments, respectively. (b) Schematic view of the Mo trimer. 
(c-e) Absorption spectra of Zn$_2$Mo$_3$O$_8$, FeZnMo$_3$O$_8$, and Fe$_2$Mo$_3$O$_8$ measured at room temperature. 
(f) A Tauc plot reveals the direct band gap of Zn$_2$Mo$_3$O$_8$ and the Fe substituted analogs. 
(g) Band gap schematic showing the impact of two different types of T and O-site substitution. The upper and lower trend lines correspond to the (Fe,Zn)$_2$Mo$_3$O$_8$ series and the $A_2$Mo$_3$O$_8$ ($A$ = Fe, Mn, Ni) materials, respectively.
%Summary of direct band gap energies of all the materials, which clearly shows Fe has strong hybridization while other elements do not.
(h,i) Calculated optical conductivity of the $A_2$Mo$_3$O$_8$ materials.
%The solid symbols on the curves inside the insets show the energies points where the optical conductivity was computed.
}
\end{minipage}
\end{figure*}

Fe$_2$Mo$_3$O$_8$ is a polar magnet with giant magnetoelectric coupling, strong Dzyaloshinski-Moriya interactions, valence bond condensation (creating a cluster magnet), and the possibility of orbitally-selective 
transitions \cite{Sheckelton2012,Mourigal2014,Wang2015,Kurumaji2015,Li2017,Chen2018,Solovyev2019,Nikolaev2021}. 
Zinc substitution, first on the tetrahedral Fe site and then on the octahedral Fe site \cite{Kurumaji2015,Streltsov2019}, is of interest for magnetic properties as well \cite{Nakayama2011,Mourigal2014,Inosov2018,Streltsov2019}.  
%As shown in Fig. \ref{Structure}, 
The structure of Fe$_2$Mo$_3$O$_8$ consists of corner-shared tetrahedral and octahedral sites separated by layers of Mo trimers [Fig. \ref{FMOBandgap}(a,b)] \cite{McCarroll1957,Ansell1966,Sheckelton2012}. 
The FeO$_4$ tetrahedron is significantly elongated and distorted, and  the FeO$_6$  octahedron is trigonally distorted as well, leading to a  C$_{3v}$ point group on both tetrahedral and octahedral Fe sites.  As a result, Fe$_2$Mo$_3$O$_8$  has no inversion symmetry. 
The system has a 61 K magnetic ordering transition to a collinear antiferromagnetic state %with uncompensated moments 
with a concomitant structural distortion  \cite{Varret1971,Czeskleba1972,Wang2015,Stanislavchuk2019,Reschke2020}. Antiferromagnetic antiphase domain boundaries have been imaged in this state \cite{Kim2018}. Fe$_2$Mo$_3$O$_8$ 
also displays a 5 T transition to the ferrimagnetic state with an extremely large magnetoelectric coefficient  \cite{Wang2015,Kurumaji2015}.  %(dM/dE = \alpha = 5700 ps/m at T = 55 K and H = 3.345 T) (\Delta P 0.08 \mu C/cm^{2} as H goes from 3.25 to 3.5 T) (Also the AFM transition is accompanied by the giant (~0.3 \mu C/cm^{2}) jump of the electric polarization)
Ni$_2$Mo$_3$O$_8$ also hosts robust magnetoelectric coupling with a field-tunable coupling mechanism \cite{Tang2021}. 
Spectroscopic highlights in Fe$_2$Mo$_3$O$_8$ include (i) nonreciprocal directional dichroism \cite{Yu2018},  phonon trends across $T_{\rm N}$ \cite{Stanislavchuk2019,Reschke2020}, and a variety of magnetic excitations in the terahertz range \cite{Csizi2020}, (ii) M\"ossbauer to confirm the 2+ charge on the iron site \cite{Varret1971,Czeskleba1972}, and (iii) studies of charge transfer via  time-dependent optical Kerr effects  \cite{Sheu2019} complemented by first principles electronic structure calculations \cite{Biwas2017,Streltsov2019,Reschke2020}.

In order to place the charge excitations on a firm foundation, we measured the optical properties of the $A_2$Mo$_3$O$_8$ family of materials (where $A$ = Fe, Ni, Mn, Zn) %and several $A$-site substituted materials 
and compared our findings with complementary 
electronic structure calculations. 
We show that the 1.7 eV gap in Zn$_2$Mo$_3$O$_8$ %(as well as the Ni and Mn analogs) 
is determined by the charge excitations of the Mo trimer. Replacing Zn on the octahedral site with Fe yields FeZnMo$_3$O$_8$. This system has a substantially reduced and renormalized  gap  determined by Fe-O hybridized bands %, and is well described as a renormalized band gap, which 
that appear due to the  periodic lattice potential. 
Further substitution yields  Fe$_2$Mo$_3$O$_8$ which has both trigonally-distorted octahedral and tetrahedral sites occupied by Fe atoms, and the gap is further reduced to 1.0 eV. 
Here, the charge gap is more complex to describe because it has mixed band and Mott features. 
In other words, some orbitals hybridize strongly with oxygen and form very narrow bands whereas other orbitals exhibit real space localization and are Mott insulating. Mixed band and  Mott gaps are commonly called orbitally- or site-selective Mott states~\cite{Chen2020,Pascut2020,Lichtenstein2001,Anisimov2002,deMedici2005,Haule2017}. 
What distinguishes Fe$_2$Mo$_3$O$_8$ from other orbitally-selective Mott systems is the narrow many-body resonance 
emanating from the edge of the flat valence band. %This resonance has strong many-body character. 
The Kondo effect is, of course, normally studied in metals. In this work, we show that the Kondo effect can also appear in  mixed metal semiconductors. % with both Mott and band gaps.
In addition, the gap in Fe$_2$Mo$_3$O$_8$ is %thus composed  of heavily mixed charge transfer excitations that are 
sensitive to magnetic ordering at 61 K due to the heavily mixed character of the charge excitations. %
Moreover, the $d$-to-$d$ excitations on the distorted octahedral Fe site are vibronically activated, and  spin-orbit related features ride on top of the distorted tetrahedral on-site excitations below the magnetic ordering transition. We discuss these findings in terms of band-Mott mixing in 3- and 4$d$-containing  quantum cluster magnets.

%%%%%%%%%%%%%%%%%%%%%%%%%%%%%%%%%%%%%%%%%%%%%%%%%%%%%%%%%%%%%%
\section*{Methods}
%%%%%%%%%%%%%%%%%%%%%%%%%%%%%%%%%%%%%%%%%%%%%%%%%%%%%%%%%%%%%%

High quality single crystals of Fe$_2$Mo$_3$O$_8$, the Zn-substituted analogs FeZnMo$_3$O$_8$, and Zn$_2$Mo$_3$O$_8$, as well as Mn$_2$Mo$_3$O$_8$ and Ni$_2$Mo$_3$O$_8$  were grown by chemical vapor transport as discussed previously \cite{Wang2015}.  Special care was taken to assure the stoichiometry of FeZnMo$_3$O$_8$.
Crystals were polished to control optical density and expose the hexagonal face. 
A Bruker 55 Fourier transform infrared spectrometer equipped with a microscope attachment was used to measure transmittance over the 0.41 - 2.0 eV energy range. Absorption was calculated as $\alpha(E)= -\frac{1}{d}{\rm ln}({\mathcal{T}}(E$)), where ${\mathcal{T}}$($E$) is the  transmittance and \emph{d} is the thickness. Performing these measurements in transmittance rather than reflectance avoids light leakage problems. Temperature was controlled by an open-flow cryostat.

For theoretical calculations we used the density functional theory (DFT) as implemented in WIEN2k \cite{Blaha2019} and a charge-self-consistent dynamical mean field theory (DMFT) as implemented in the eDMFT code~\cite{Haule,Haule2018}. At the DFT level, we
used the generalized gradient approximation Perdew-Burke-Ernzerhof (GGA-PBE) functional \cite{Perdew1996}, with RKmax = 7.0 and 312 k-points in the irreducible part of the 1$^{st}$ Brillouin zone. At the eDMFT level, we used the fully rotationally invariant Coulomb interaction, a nominal double counting scheme \cite{Haule2015}, with the $d$-orbital occupations for double counting corrections for Mn, Fe, Co and Ni set to be 5, 6, 7 and 8, respectively. The temperature is fixed at $500\,$K. To define the DMFT projector, we used quasi-atomic
orbitals by projecting bands in a large hybridization window (-10 to +10 eV) with respect to the Fermi level, in
which partially screened Coulomb interactions have values of
U = 10 eV and J$_H$ = 1 eV in Mn, Fe, Co and Ni ions. 
In order to solve the auxiliary quantum impurity problem, a continuous-time quantum Monte Carlo method in the hybridization-expansion (CT-HYB) was used~\cite{Haule2007}, where the five $d$ orbitals for the Mn, Fe, Co and Ni ions (grouped according to the local C$_{3v}$ point group symmetry) were chosen as our correlated subspace in a single-site DMFT approximation. For the CT-HYB calculations, up to 10$^8$ Monte Carlo steps were employed for each Monte Carlo run. The  self-energy  on  the  real  axis  was  obtained  using  the  analytical  continuation maximum entropy method for the local cumulant as explained in \cite{ME_haule}. During the calculation, the position of the chemical potential was kept fixed within the gap.
The experimental crystal structures used for our computations \cite{Stanislavchuk2019} as well as details of our calculations are given in the Supplementary information \cite{Supp}.

%%%%%%%%%%%%%%%%%%%%%%%%%%%%%%%%%%%%%%%%%%%%%%%%%%%%%%%%%%%%%%
\section*{Results and Discussion}
%%%%%%%%%%%%%%%%%%%%%%%%%%%%%%%%%%%%%%%%%%%%%%%%%%%%%%%%%%%%%%

\subsection*{Optical response of Fe$_2$Mo$_3$O$_8$ and the $A$-substituted analogs ($A$ = Zn, Mn, Ni)}

Figure \ref{FMOBandgap}(c-e) summarizes the optical properties of the (Fe,Zn)Mo$_3$O$_8$ family of materials. The absorption spectrum of the parent compound, Zn$_2$Mo$_3$O$_8$, is low and flat in the near infrared, rising on approach to  the O 2$p$ $\rightarrow$ Mo 3$d$ charge transfer excitation. The  direct band gap is 1.75 eV.  Because Zn$^{2+}$ has a $d^{10}$ configuration, there are no $d$-to-$d$ on-site excitations.  Zn$_2$Mo$_3$O$_8$ therefore provides an opportunity to study how the Mo trimer interacts with oxygen in isolation. At the same time, it is an important %framework 
scaffold upon which additional complexity can be built.

Sequential $A$-site substitution of Fe, first into the distorted octahedral site in FeZnMo$_3$O$_8$, here denoted as Fe(O), and then into the distorted tetrahedral site in Fe$_2$Mo$_3$O$_8$, henceforth Fe(T), lowers the charge gap significantly [Fig.~\ref{FMOBandgap}(f)]. We find direct gaps of 1.2 and 1.0 eV for FeZnMo$_3$O$_8$ and  Fe$_2$Mo$_3$O$_8$, respectively.  
The gap values were determined from Tauc plots  of 
(${\alpha}{\cdot}E$)$^2$ vs. energy \cite{Pankove2010}.

\begin{figure*}[tbh]
\begin{minipage}{2.2in}
\caption{(a, d) Tauc plot of (${\alpha}{\cdot}E$)$^2$ vs. energy  for Fe$_2$Mo$_3$O$_8$ and a plot of band gap vs. temperature  with a fit to the Varshni model. (b, e) Close-up view of the Fe$^{2+}$ $d$-to-$d$ on-site excitation on the tetrahedral site showing the fine structure that develops due to spin-orbit coupling below the 61 K magnetic ordering transition. (c, f) Close-up view of the Fe$^{2+}$ $d$-to-$d$ on-site excitation on the octahedral site and oscillator strength vs. temperature along with a fit to a modified vibronic coupling model \cite{Ballhausen1962,Stoneham2001,ONeal2017}.\label{Temperature}}
\end{minipage}
\begin{minipage}{4.4in}
\includegraphics[width=4.4in]{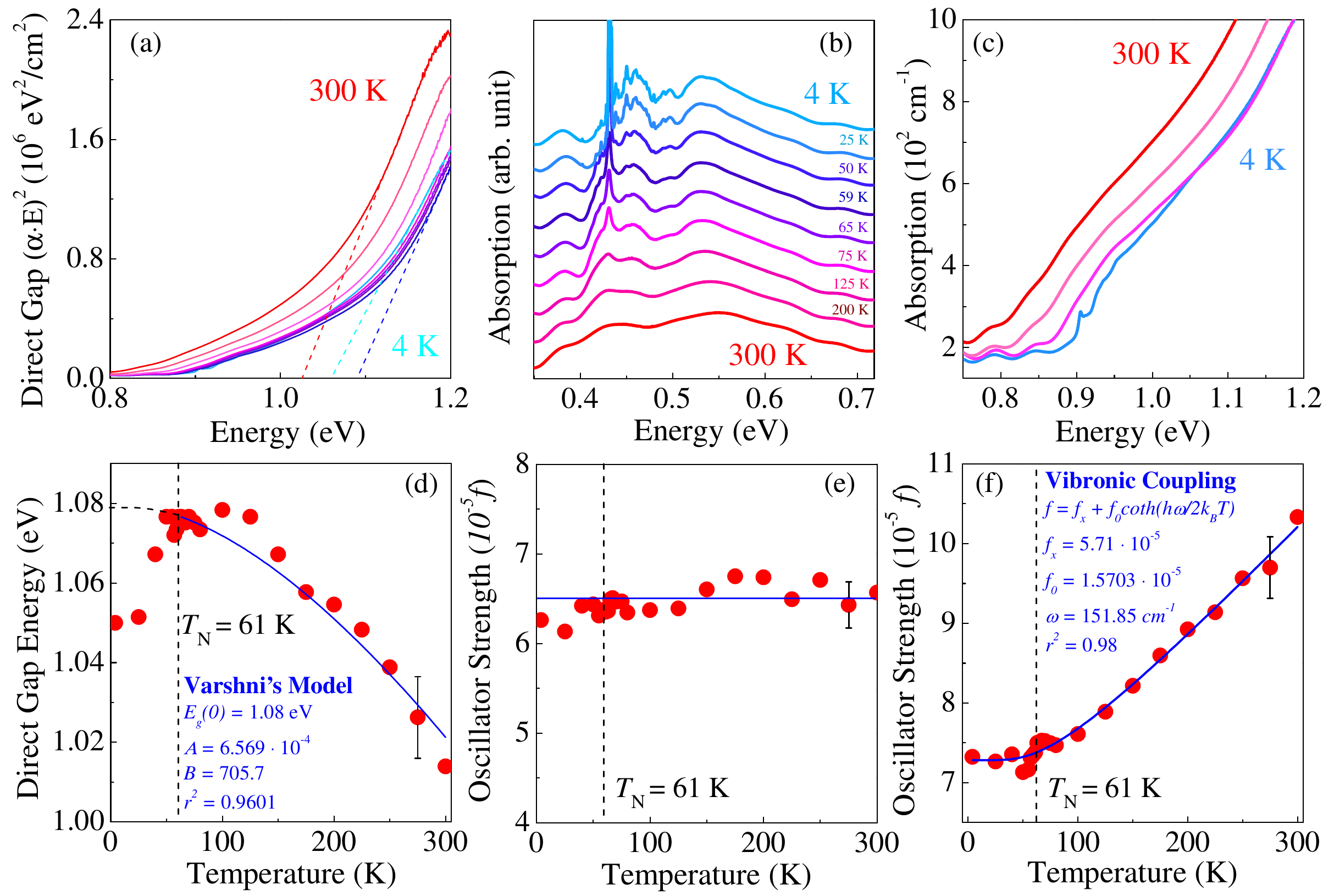}
\end{minipage}
\end{figure*}

Figure~\ref{FMOBandgap}(d) displays the absorption of  FeZnMo$_3$O$_8$. The charge excitations across the gap consist of mixed O 2$p$ + Mo 3$d$ + Fe(O) 3$d$  transitions. The lowest energy excitation across this gap comes from Fe(O) hybridizing with Mo-O trimers. The 1.2 eV gap is substantially lower in energy than the fundamental Mo-O band gap, which theoretically remains roughly equal to that in Zn$_2$Mo$_3$O$_8$. As Fe$^{2+}$ populates the octahedral site in FeZnMo$_3$O$_8$, an on-site $d$-to-$d$ transition arises near 0.95 eV. It overlaps strongly with the leading edge of the charge transfer band and is activated by vibronic coupling [Fig. \ref{Temperature}(c,f)] \cite{Ballhausen1962,Stoneham2001,ONeal2017}. Notice that absorption is low and flat near 0.5 eV - a sign of crystal quality and stoichiometry. Once the Fe(T) site is populated as well (as in Fe$_2$Mo$_3$O$_8$), the gap is reduced further, and two
 different types of $d$-to-$d$ on-site excitations are identified inside the charge gap. As shown in Fig.~\ref{FMOBandgap}(e), Fe on the trigonally-distorted tetrahedral site contributes additional atomic-like excitations centered at 0.5\,eV. While oscillator strength is fully conserved as a function of temperature [Fig. \ref{Temperature}(e)], a great deal of fine structure due to spin-orbit coupling rides on top of the band  below the 61 K magnetic ordering transition [Fig. \ref{Temperature}(b)]. A similar response develops in Fe$^{2+}$:ZnSe \cite{Evans2017}. The behavior of the Fe$^{2+}$ on-site $d$-to-$d$ excitations in Fe$_2$Mo$_3$O$_8$ is summarized in Fig. \ref{Temperature}(b,c,e,f) and further discussed in Supplementary information \cite{Supp}. The full sequence of gap values
is shown schematically in Fig.~\ref{FMOBandgap}(g).

%\subsection*{Optical properties of the $A_2$Mo$_3$O$_8$ materials ($A$ = Fe, Mn, Ni)}

To test the influence of $A$-site substitution on the band gap and strength of the metal to MoO-trimer hybridization, we measured the optical properties of the Mn and Ni analogs of Fe$_2$Mo$_3$O$_8$ [Fig. S2, Supplementary information] \cite{Supp}.
Mn$_2$Mo$_3$O$_8$ and Ni$_2$Mo$_3$O$_8$ have  charge gaps of 1.65 and 1.7 eV, respectively - very similar to that of the Zn end member.
This result is attributable to 3$d$ orbital filling and character. The  Mn system has a half-filled $d$-manifold, % ($d^5$) - 
which corresponds to a high spin Mott state with a large gap. % and the $d$ orbitals are in the high spin Mott state with a large gap.
The $d^8$ configuration in the Ni analog also has two holes and is thus in the large gap Mott insulating state. As a result, there is little mixing %minimal interaction
between the metal center and Mo-O trimer, hence these transition metals do not play an active role in determining the low-energy excitations across the gap. We mention in passing that the Mn and Ni compounds have on-site $d$-to-$d$ excitations as well [Fig. S2, Supplementary information] \cite{Supp}.

On the other hand, Fe$_2$Mo$_3$O$_8$ has both Mott-type and band-insulating orbitals, which strongly hybridize with the Mo trimer. These  interactions enable the Fe centers to control the low energy physics as discussed below. Moreover, we find that the band gap of Fe$_2$Mo$_3$O$_8$ is sensitive to the 61 K magnetic ordering transition and the associated structural distortion [Fig. \ref{Temperature}(a,d)].
This is different from the other $A_2$Mo$_3$O$_8$ compounds, where the temperature dependence of the band gaps are in excellent overall agreement with the Varshni model \cite{Sarswat2012,ODonnell1991} implying no (or extremely subtle) structural aspects to the magnetic ordering transitions. That the band gap decreases across $T_{\rm N}$ is due to coupling of charge, structure, and magnetism and the flat bands emanating from the trigonally-distorted tetrahedral Fe site. % as described below. 

\begin{figure*}[tbh]
\begin{minipage}{7.0in}
\includegraphics[width = 7.0in]{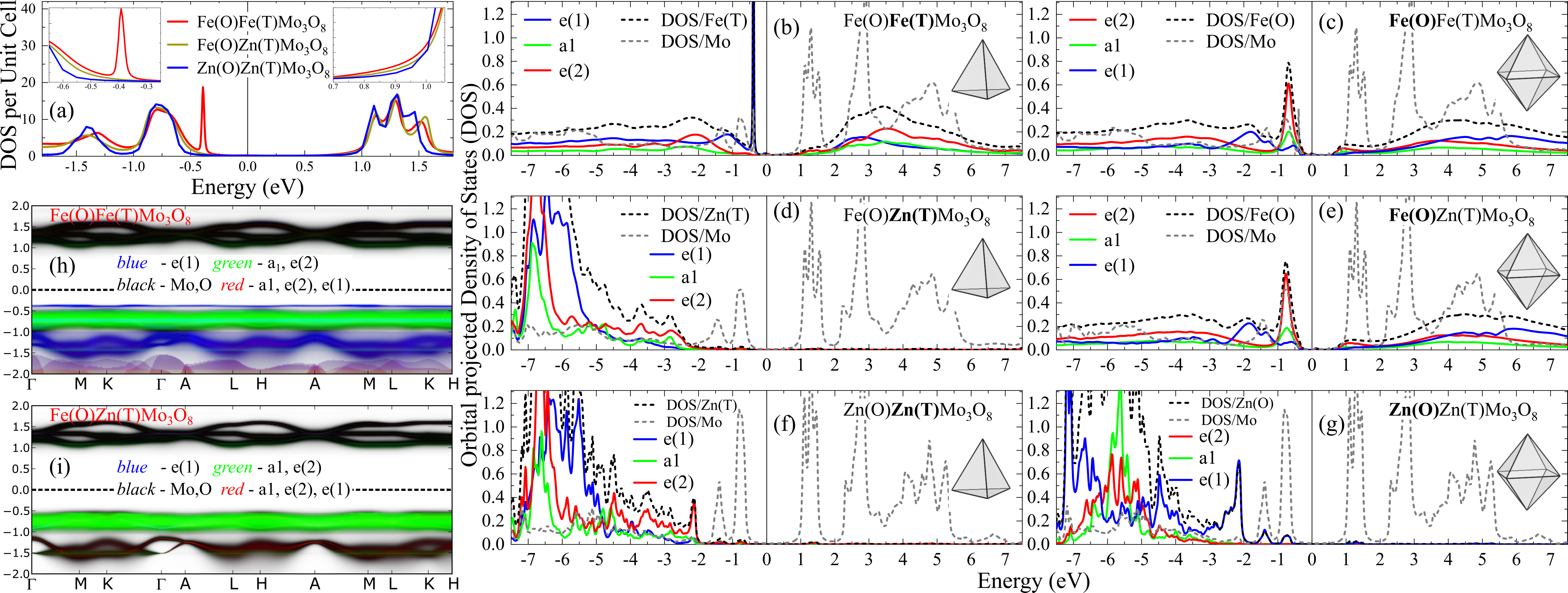}
\end{minipage}
\begin{minipage}{7.0in}
\caption{\label{DOS} 
Density of states (DOS) for $A_2$Mo$_3$O$_8$ ($A$ = Fe and Zn): (a) total DOS; (b-g) atom- and orbital-projected DOS. % (h, i) schematic view of the orbitals. 
The (T) and (O) symbols refer to  trigonally-distorted tetrahedral and octahedral environments. 
%For example, Fe(T) and Fe(O) are the labels for Fe$^{2+}$ in a  distorted tetrahedral and octahedral environment, respectively.  
The vertical solid lines are placed at zero chemical potential.  The schematic insets of gray tetrahedra$/$octahedra are guides to the eye pointing the reader to the electronic states of the transition metal ions in the corresponding environment. (h, i) Orbital projected spectral functions for Fe$_2$Mo$_3$O$_8$ and FeZnMo$_3$O$_8$ (blue - $e$(1) tetrahedra; green - $a_1$ and $e$(2) octahedra; red - $a_1$, $e$(2) tetrahedra and $e$(1) octahedra)}.
\end{minipage}
\end{figure*}

\begin{figure*}[tbh]
\begin{minipage}{7.0in}
\includegraphics[width = 7.0in]{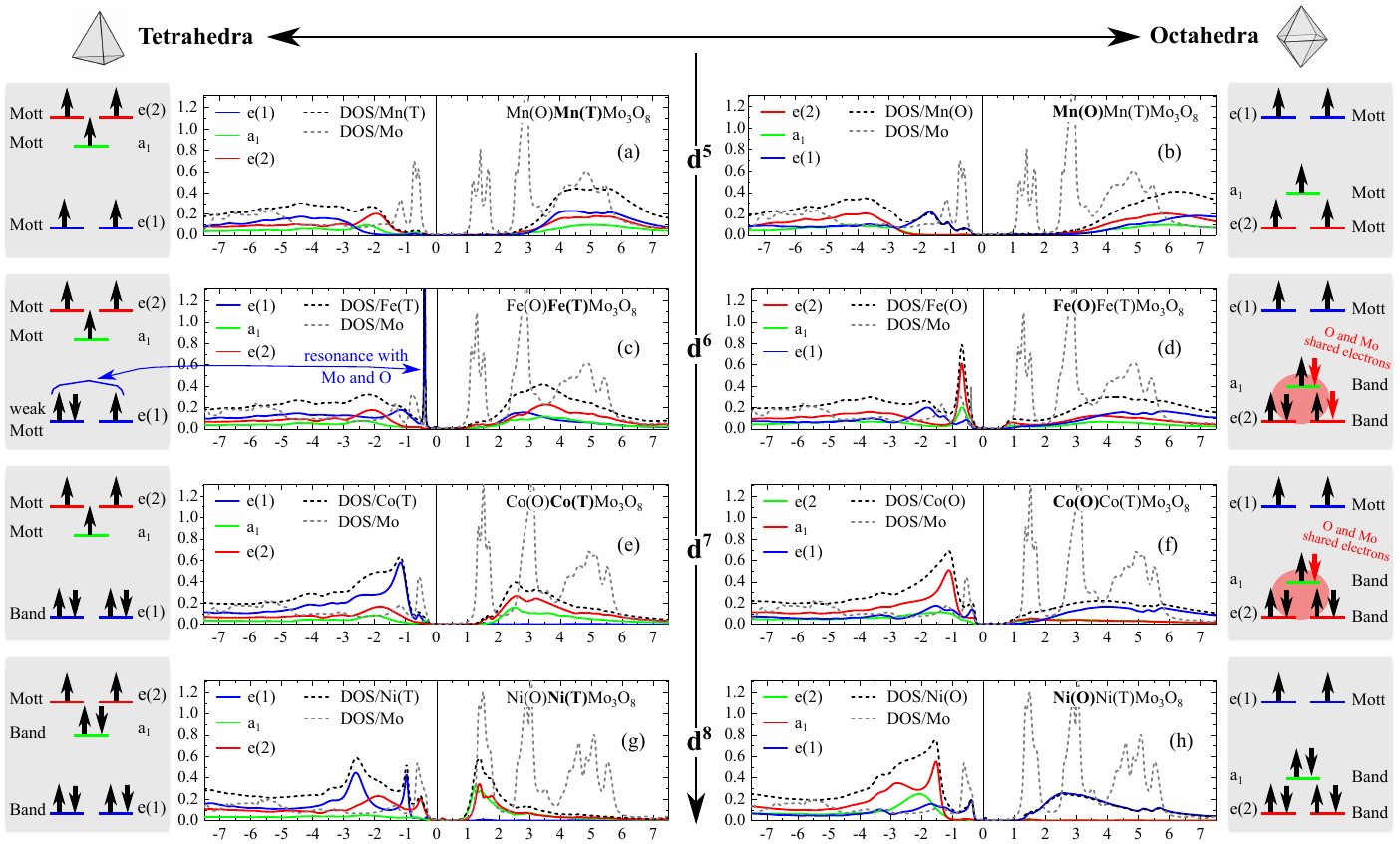}
\end{minipage}
\begin{minipage}{7.0in}
\caption{\label{DOS_orbitals}
Atom- and orbital-projected density of states (DOS) together with a schematic view of the orbital occupation and character for $A_2$Mo$_3$O$_8$ ($A$ = Mn, Fe, Co and Zn). 
%The (T) and (O) symbols refers to the tetrahedra and octahedra environments, for example Mn(O) is the label for the Mn ion inside the octahedra environment and e$_g$(T) is the label for the e$_g$ orbital of the corresponding TM(T) ion inside the tetrahedra environment. 
Each panel shows a schematic view of the  orbital occupation (left side) and DOS (right side). Panels (a, c, e, g) refer to ions in trigonally-distorted  tetrahedral environments ($A$(T)), whereas panels (b, d, f, h) refer to ions in trigonally-distorted octahedral environments ($A$(O)). 
%The vertical solid lines placed at zero represents the chemical potential. The schematic insets of gray tetrahedra$/$octahedra are guide to the eyes pointing the reader to the electronic states of the TM ions inside that corresponding environment.
}
\end{minipage}
\end{figure*}

\subsection{Strong hybridization, resonance, and interaction with the Mo trimer}

Figure~\ref{FMOBandgap}(h) displays the theoretical optical conductivity of Zn$_2$Mo$_3$O$_8$, FeZnMo$_3$O$_8$ and Fe$_2$Mo$_3$O$_8$ computed using a combination of Density Functional Theory and embedded Dynamical Mean Field Theory (DFT + eDFMT) methods \cite{Haule,Haule2018}.
Here, we find the same trend of decreasing charge gap with Fe substitution. In Zn$_2$Mo$_3$O$_8$, the size of the optical gap  is $\approx$1.7$\,$eV which decreases to approximately 1.5 and 1.4 eV in 
FeZnMo$_3$O$_8$ and Fe$_2$Mo$_3$O$_8$, respectively.
Figure~\ref{FMOBandgap}(i) compares the theoretical optical conductivity of the Mn, Ni and Co analogs. We find that the predicted gap is larger in all of these compounds (near $1.55\,$eV) as compared to Fe$_2$Mo$_3$O$_8$. The edge of the gap is very smooth and temperature smeared. This is quite different from  Fe$_2$Mo$_3$O$_8$ where  the gap is smaller with an additional peak at the onset.

To better understand what determines the low energy excitations and character of the gap in this class of compounds, we calculated the local density of states for the full set of $A_2$Mo$_3$O$_8$ materials [Fig.~\ref{DOS}]. While the optical gap in general is different than the gap of the single-particle excitations measured by the local density of states,  the two are very similar in these compounds. This is because the band gap is direct in Zn$_2$Mo$_3$O$_8$, and the hybridized bands in the Fe-containing  compounds are extremely narrow. Hence momentum-conserving excitations have essentially the same gap size as the finite momentum single particle excitations.
The insets show that the position of the conduction band in FeZnMo$_3$O$_8$ and Fe$_2$Mo$_3$O$_8$ decreases slightly as compared to Zn$_2$Mo$_3$O$_8$, although the change is small. Most of the action is in the valence bands, where the FeZn band edge moves considerably upward. In Fe$_2$Mo$_3$O$_8$, a very narrow many-body excitation forms at the onset of the gap. This is the origin of the first peak in the optical conductivity [Fig.~\ref{FMOBandgap}(h,i)] and the reason that the gap is drawn strongly downward in this system.

Figure~\ref{DOS}(b-g) displays the projected density of states (DOS)  per transition metal center and per orbital in the distorted tetrahedral (T) and octahedral (O) environments. The local point group symmetry around each iron center is C$_{3v}$. Therefore, the $e$ and $t_2$ levels at the distorted tetrahedral site break into $e$(1), $a_1$, and $e$(2) orbitals. Similarly symmetry at the trigonally-distorted Fe octahedral site breaks $t_{2g}$ and $e_g$ into $e$(2), $a_1$, and $e$(1) states.  These energy levels are  shown in Fig. \ref{DOS_orbitals}, and  the symmetry breaking is diagrammed in Fig. S1, Supplementary Information \cite{Supp}.

Getting back to Figure~\ref{DOS}(b-g), we notice that the Mo states (dotted grey line) are very similar across this entire family of materials. The low-energy excitations are, however, not on the Mo site when Fe is present. Figure~\ref{DOS}(b) reveals that the sharp many-body resonance near $-0.4\,$eV emanates primarily from the Fe center in the trigonally-distorted tetrahedral environment, $e$(1). Panel (c) shows that a broader, but still reasonably sharp excitation around $-0.8\,$eV arises from the doubly-degenerate $e$(2) state on the distorted octahedral site. Since both of these fairly sharp excitations come from band formation via hybridization, the Mo partial density of states also has a small peak at the same energy (Figs. S19 and S20). This demonstrates the quasi-particle nature of these peaks, which are Kondo-like and come from local spin screening on the aforementioned $e$(1) 
and $e$(2) distorted tetrahedral and octahedral %(O) 
orbitals, respectively.

What is exciting about this finding is that screening and many body Kondo peak formation \cite{Yee2010,Hewson1993} are normally expected in a metal - not an insulator. 
%This screening instead takes place in an insulator.  
The possibility of a Kondo resonance in a semiconductor like Fe$_2$Mo$_3$O$_8$ is potentially quite interesting, opening the door to deeper exploration of the Kondo effect in a significantly wider variety of materials.   %rather than a metal - which is normally 
%Of course here, the screening is found in an insulator rather than a metal - as 
%expected for .
While broader peaks like that at $-0.8\,$eV are not uncommon in transition metal compounds and appear for example in monoxides~\cite{Mandal2019,Mandal2019_1},
% and are usually called generalized Zhang-Rice singlets~\cite{Zhang1988,Eskes1988}, 
the very narrow resonance emanating from the  $e$(1) orbitals on the distorted tetrahedral site is unique to Fe$_2$Mo$_3$O$_8$. We assign it as an analog of the Zhang-Rice singlet in cuprates,~\cite{Zhang1988,Eskes1988} arising  due to screening of the spin $1/2$ hole on the trigonally-distorted tetrahedral  Fe $e$(1) sites in the Mott insulating state. This characteristic peak appears with very well defined energy. It also mixes strongly with the Mo-O trimers (see supplement Fig. S5).

%\textcolor{red}{While Fe$_2$Mo$_3$O$_8$ displays sharp peaks in the valence band spectra of both the trigonally-distorted octahedral $e$(2) and tetrahedral $e$(1) orbitals [Fig. \ref{DOS}(b,c)], FeZnMo$_3$O$_8$  has a Zn atom in the \textcolor{blue}{distorted} tetrahedral site.} Here, Zn$^{2+}$ is in the $d^{10}$ configuration and therefore inert because the first features are below $-2\,$eV. The distorted Fe octahedral site still shows a sharp valence band peak in the $e$(2)
%orbitals. It is this peak on the edge of the valence band that determines the size of the gap in FeZnMo$_3$O$_8$. It is also the physical reason for the reduction of the gap as compared to Zn$_2$Mo$_3$O$_8$. Finally, panels (f) and (g) show the Zn-projected density of states in Zn$_2$Mo$_3$O$_8$. The Zn states are occupied and far below the Fermi energy. In this case, the Mo trimers determine the low-energy excitations and the gap size. Because the density of states are momentum-averaged, we also show the spectral functions [Fig. \ref{DOS}(h,i)].

%\newpage 

In order to test these predictions, we compare  the theoretical optical conductivity [Fig. \ref{FMOBandgap}(h)] to the measured absorption spectrum of Fe$_2$Mo$_3$O$_8$ [Fig. \ref{FMOBandgap}(f)]. Overall, the calculated optical conductivity and the measured absorption spectrum are very consistent - although the %predicted experimental 
features are not as well-defined as we might prefer. As a reminder, the spectral functions are very flat on the valence edge due to Fe occupation on the distorted tetrahedral site. This causes a sharp many-body resonance %on the edge of the valence band 
in the density of states [Fig \ref{DOS}(b)] which manifests as a small peak on the leading edge of the theoretical optical conductivity. The contribution from the distorted  octahedral site is similar but less pronounced. Our calculations therefore predict that the many-body resonance on the valence band edge [Fig \ref{DOS}(b,c)] should lower the gap.   
This is exactly what we find. Obviously this structure is strongly broadened in the experimental result, but the presence of these states causes the gap to be  drawn noticeably downward in this system [Fig.~\ref{FMOBandgap}(f)]. Similar reasoning applies to FeZnMo$_3$O$_8$, although only the distorted octahedral site is operative. Because the density of states are momentum-averaged, we also show the spectral functions [Fig. \ref{DOS}(h,i)].

\subsection*{Structure-property relations in the metal-substituted analogs}

Figure \ref{DOS_orbitals} compares the density of states in Fe$_2$Mo$_3$O$_8$ with several other transition metal analogs which have orbital filling between $d^5$ and $d^8$ - namely the Mn, Co and Ni analogs. We also show a schematic view of the orbital occupation for both the distorted tetrahedral and octahedral sites (left and right columns, respectively). The gap in each orbital can originate from the band structure due to the periodic potential (band gap) or from Mott localization of electrons on a given $A$ site, which we denote as a Mott gap. In principle, we can distinguish between the two because the single-particle spectral function is modified from the DFT bands by self-energy effects. We expect the spectral function to be either zero or finite in the band gap case; it should diverge inside the gap for the Mott case  \cite{Pavarini2017,Demchenko2004,Kotliar2006}. Here, the electron state can no longer be described within the band picture. 

%At the valence edge of the gap, however, hybridization can still provide an efficient screening channel, even though no itinerant state is present at the Fermi level. In this case a sharp resonance or a quasi-particle multiplet can appear at the gap edge~\cite{Yee2010} which is temperature dependent and sensitive to small perturbations. We have identified such a  many-body resonance in the $e_g$(T)\textcolor{blue}{-like} orbital in
%Fe$_2$Mo$_3$O$_8$ above. %Interestingly, a similar structure appears in NiPS$_3$ \cite{Kang2020}. 

The top panels of Fig.~\ref{DOS_orbitals} show calculations for Mn$_2$Mo$_3$O$_8$ in which the electrons are in the high-spin $d^5$ configuration. The Mott gap opens in all orbitals on both tetrahedral and octahedra sites, and it is much larger than the band gap of Mo-trimer. Consequently the charge gap and low energy excitations in the Mn and Zn compounds are determined by the same Mo-trimer states. This is why they appear so similar.

Fe$_2$Mo$_3$O$_8$ is the most interesting of the series showing a complex interplay of band gaps, Mott gaps, and quasi-particle multiplets. The $a_1$ and $e$(2) 
states on the distorted tetrahedra are Mott insulating with a large gap. As discussed above, the doubly degenerate $e$(1)
orbitals contain one hole, which is equally distributed among the two orbitals, and the Mott mechanism opens the gap - even though the self-energy pole is less strong and the gap smaller than in the $a_1$ and $e$(2)
states. Moreover, at the edge of the gap, strong hybridization, directly computable by the DMFT hybridization function, shows a very strong and narrow peak due to many body screening effects. With this mechanism, the spin-$1/2$ emanating from the $e$(1) 
 orbitals on the tetrahedral site is screened by the Mo-trimer electrons - a mechanism that is analogous to the Zhang-Rice state in cuprates~\cite{Zhang1988}. Note that this is different from the sharp peak due to the valence band edge in transition metal monoxides \cite{Mandal2019,Mandal2019_1}. The latter appears in the antiferromagnetic state where all gaps are band-like in nature and spin states are split by a Zeeman field. As a consequence, many-body screening of spin is not possible. Here, spin preserves SU(2) symmetry, and the resonance screens the local spin on Fe. In addition, the narrow peak due to many body screening effects disappears in the antiferromagnetic ground state [Fig. S19, Supplementary information] \cite{Supp}.
The distorted octahedral site in the Fe $d^6$ state contains a combination of a Mott gap in the $e$(1)
states and a  band gap in the $a_1$ and $e$(2)
states. This unusual combination is a so-called orbitally-selective Mott state~\cite{Anisimov2002,deMedici2005}. Note that the sharp peak at the valence band edge appears as well, even though it is not as sharp as the resonance from the distorted tetrahedral Fe $e$(1) 
site. Its nature is different, as it appears in band-insulating $a_1$ and $e$(2)
orbitals similar to the valence band edge  discussed in transition metal monoxides~\cite{Mandal2019_1,Mandal2019}. Note that hybridization and band formation with oxygen and Mo electrons are needed to open the band gap in the octahedral Fe $a_1$ and $e$(2) 
states. This is because the latter contains only four electrons in three nearly degenerate $a_1$ and $e$(2)
orbitals.

Next we discuss the Co analog, which is in the $d^7$ configuration. In this case, the $e$(1) %$e_g$(T)\textcolor{blue}{-like} 
orbitals on the tetrahedral site are fully filled, and the $a_1$ and $e$(2) %$t_{2g}$(T)\textcolor{blue}{-like} 
orbitals are in the high-spin Mott insulating state with a large gap - larger than the  Mo-trimer gap. On the octahedral site, the $e$(1) %\textcolor{blue}{quasi-}$e_g$(O) 
orbitals are Mott-insulating with a large gap, and the $a_1$ and $e$(2) %\textcolor{blue}{quasi-}$t_{2g}$(O) 
states are band-insulating in which Mo and oxygen provide one electron to form a covalent band with the $a_1$ %$t_{2g}$(O)\textcolor{blue}{-like} 
electrons. We note that no sharp low-energy peak is found at the valence band edge, although in principle such a peak is possible.

Figure~\ref{DOS_orbitals}(g,h) displays the partial density of states for the Ni analog with its $d^8$ configuration. In this case, the $e$(1) 
and  $a_1$ orbitals on the distored tetrahedral site
are fully filled, and the doubly-degenerate $e$(2) 
state shows a Mott gap  which is comparable to that of the Mo trimer. On the distorted octahedral site, the $e$(2) and $a_1$ %\textcolor{blue}{quasi-}$t_{2g}$(O) 
states are fully filled, and the $e$(1) %$e_g$(O)\textcolor{blue}{-like} 
states are in the half-filled Mott insulating state. This Mott gap is again comparable to the Mo-trimer gap. Hence the reduction of the gap as 
compared to the Zn analog is minimal.

\section*{Summary and outlook}

To summarize, we measured the optical properties of Fe$_2$Mo$_3$O$_8$ and compared our findings with 
first principles electronic structure calculations. 
We find a 1.1 eV direct gap composed  of heavily mixed charge-transfer excitations that is sensitive to magnetic ordering at 61 K, 
vibronic coupling that activates on-site $d$-to-$d$ excitations on the distorted octahedral Fe site, and spin-orbit related features riding on top of the   $d$-to-$d$ excitation on the distorted tetrahedral Fe site below the magnetic ordering temperature.
The Kondo effect is, of course, usually studied in metals. Here, we show that it can also appear in a semiconductor that has both Mott and band gaps. Similar to the metallic Kondo effect, the orbitals with Mott-like gaps develop a many-body excitation near the valence edge. This draws the gap downward in energy (from 1.7 eV in Zn$_2$Mo$_3$O$_8$  $\rightarrow$ 1 eV in Fe$_2$Mo$_3$O$_8$) and screens the magnetic moment. This discovery opens the door to deeper exploration of the Kondo effect in semiconductors. 
Fe$_2$Mo$_3$O$_8$ is also a superb platform for unraveling structure-property relationships. What differentiates Fe$_2$Mo$_3$O$_8$  from the Zn, Mn, Co, and Ni members of this series is the band-Mott mixing, the Zhang-Rice resonance,  
%in this $d^6$ system
and how the gap is hybridized.
Taken together, these findings enhance our understanding of charge transfer in quantum cluster magnets and advance the use of this powerful scaffold in new types of %batteries. 
charge storage devices.

\section*{Acknowledgements}

Research at the University of Tennessee and Rutgers University is supported by the NSF-DMREF program (DMR-1629079 and DMR-1629059). G.L.P.'s work was supported by a grant of the Romanian Ministry of Education and Research, CNCS - UEFISCDI, project number PN-III-P1-1.1-TE-2019-1767, within PNCDI III. Access to the x-ray facilities at the Research Complex, Rutherford Appleton Laboratory is gratefully acknowledged.

%\nocite{Villars2016,CrysAlisPro,Petricek,Clark1995,Becker1974,Momma2011,Inkscape,Abe2010,Cuny2009}

\bibliographystyle{apsrev4-1}
\bibliography{citation}

\end{document}